\documentclass[showpacs,aps,twocolumn]{revtex4}
\usepackage{graphicx}

\def\ket{\rangle}
\usepackage{graphicx}
\usepackage{amsmath}

\begin{document}

\title{ Multi-step quantum secure direct communication using multi-particle Green-Horne-Zeilinger state}
\author{Chuan Wang$^{1}$, Fu Guo Deng$^{3}$ and Gui Lu Long$^{1,2}$}
\affiliation{$^1$ Department of Physics and Key Laboratory For
Quantum Information and Measurements, Tsinghua University, Beijing
100084, People's Republic of China \\
$^2$ Key Laboratory for Atomic and Molecular Nanosciences,
Tsinghua University, Beijing 100084, People's Republic of China\\
$^3$ The Key Laboratory of Beam Technology and Materials
Modification of Ministry of Education, and Institute of Low Energy
Nuclear Physics, Beijing Normal University, Beijing 100875,
People's Republic of China}
\date{\today }

\begin{abstract}
A multi-step quantum secure direct communication protocol using
blocks of multi-particle maximally entangled state is proposed. In
this protocol, the particles in a Green-Horne-Zeilinger state are
sent from Alice to Bob in batches in several steps. It has the
advantage of high efficiency and high source capacity.
\end{abstract}

\pacs{03.67.Dd,03.67.Hk,03.67.-a} \maketitle

Quantum communication becomes one of the most important applications
of quantum mechanics today. Quantum key distribution(QKD) is one of
the most mature techniques, providing a secure way for creating a
private key with which two authorized parties, Alice and Bob, can
realize secure communication. QKD has attracted a widespread
attention and progressed quickly \cite{bb84,bbm92,b92,bw,gisin,guo}
over the past two decades.

Recently, a new concept in quantum cryptography, quantum secure
direct communication (QSDC) was proposed. Upto date, QSDC has been
studied by many groups
\cite{beige,bf,long1,long2,cai1,yan0,yan1,zhangzj,cai2}. With QSDC
Alice and Bob can exchange the secret message directly without
generating a private key first and then encrypting the secret
message and then send the ciphertext through classical
communication. The QSDC protocol originally proposed by Beige et al.
\cite{beige} is only a scheme for transmitting secret information
which can be read out with an additional classical information for
each qubit. Later, Bostr\"{o}m and Felbinger put forward a ping-pong
QSDC scheme following the idea of quantum dense coding \cite{bw}
with Einstein-Podolsky-Rosen (EPR) pairs. As pointed out by their
authors, ping-pong protocol is just a quasi-secure direct
communication protocol as it leaks some of the secret message in a
noisy channel. In Ref. \cite{long1}, some of us proposed a two-step
QSDC protocol using entangled particles. In this paper, the idea of
transmitting quantum data in blocks for the security of QSDC was
proposed.

Recently, Gao et al. \cite{Gao} designed a protocol for controlled
quantum teleportation and secure direct communication using GHZ
state.

As GHZ-state has a larger Hilbert space, protocols based on them can
provide larger source capacity. With the development of technology,
the generation and manipulation are becoming more sophisticated, and
practical applications of these highly entangled objects are
expected in the future. It is thus interesting to look for ways of
using these quantum objects in quantum key distribution.  In this
paper, we present a  QSDC protocol using maximally entangled
three-particle Green-Horne-Zeilinger(GHZ)  states. It will be shown
that this QSDC protocol appears to provide better security. In
addition, this QSDC protocol can also be equipped with quantum
privacy amplification so that it can work under realistic
environment.

We first introduce the idea of GHZ state dense coding briefly.
Dense coding using three particle entangled states has been
proposed by Lee et al. \cite{Lee}. The protocol we propose is the
application of three particle dense coding to QKD.  Suppose the
maximally entangled three-particle state is
\begin{equation}
|\Psi\rangle_{ABC}=\frac{1}{\sqrt{2}}(|000\rangle_{ABC}+|111\rangle_{ABC}).
\end{equation}
There are eight independent GHZ-states, namely
\begin{eqnarray}
  &|\Psi\rangle_{1}=\frac{1}{\sqrt{2}}(|000\rangle+|111\rangle) ,
  &|\Psi\rangle_{2}=\frac{1}{\sqrt{2}}(|000\rangle-|111\rangle) ;\nonumber\\
 & |\Psi\rangle_{3}=\frac{1}{\sqrt{2}}(|100\rangle+|011\rangle) ,
  &|\Psi\rangle_{4}=\frac{1}{\sqrt{2}}(|100\rangle-|011\rangle) ;\nonumber\\
  &|\Psi\rangle_{5}=\frac{1}{\sqrt{2}}(|010\rangle+|101\rangle) ,
  &|\Psi\rangle_{6}=\frac{1}{\sqrt{2}}(|010\rangle-|101\rangle) ;\nonumber\\
 & |\Psi\rangle_{7}=\frac{1}{\sqrt{2}}(|110\rangle+|001\rangle) ,
 & |\Psi\rangle_{8}=\frac{1}{\sqrt{2}}(|110\rangle-|001\rangle).\label{e2}
\end{eqnarray}

By performing single-particle unitary operations on any two of the
three particles, one can change from one GHZ-state to another. The
unitary operations are the product of the Pauli and identity
matrices: $I, \sigma_x, i\sigma_y, \sigma_z$. Though there are
altogether 16 such operations, only half of them can generate
distinguishable states. For our purpose, we choose the following
eight operations
\begin{eqnarray}
 & U_1=\sigma_z\otimes\sigma_z, &U_2=I\otimes\sigma_z; \nonumber\\
  &U_3=i\sigma_y\otimes\sigma_z,&U_4=\sigma_x\otimes\sigma_z; \nonumber\\
  &U_5=I\otimes\sigma_x, &U_6=\sigma_z\otimes\sigma_x;\nonumber
  \\
  &
  U_7=\sigma_x\otimes\sigma_x,&U_8=i\sigma_y\otimes\sigma_x,\label{e3}
  \end{eqnarray}
to  realize the following transformation,
\begin{equation}
U_k|\Psi\rangle_1=|\Psi\rangle_{k},k=1,2,...,8.\label{e4}
\end{equation}
Starting from the state $|\Psi\rangle_{1}$,  one can match each
unitary operation with a GHZ-state uniquely.

We now describe the multi-step quantum secure direct communication
using GHZ state in detail. First Alice and Bob make an agreement
that each of the states $|\Psi\rangle_{k}$ represents a three bits
binary number, namely $|\Psi\rangle_1\Longrightarrow000$,
$|\Psi\rangle_2\Longrightarrow001$,
$|\Psi\rangle_3\Longrightarrow010$,
$|\Psi\rangle_4\Longrightarrow011$,
$|\Psi\rangle_5\Longrightarrow100$,
$|\Psi\rangle_6\Longrightarrow101$,
$|\Psi\rangle_7\Longrightarrow110$,
$|\Psi\rangle_8\Longrightarrow111$ respectively. Then the specific
steps in the QSDC protocol is

Step 1: Alice produces a sequence of $N$ GHZ states. Each
GHZ-state is in state
\begin{equation}
|\Psi\rangle_{ABC}=\frac{1}{\sqrt{2}}(|000\rangle_{ABC}+|111\rangle_{ABC}).
\end{equation}
We represent the sequence of $N$ GHZ-states as
 [$P_1(A)P_1(B)P_1(C)$,
$P_2(A)P_2(B)P_2(C)$, ..., $P_N(A)P_N(B)P_N(C)$], where the numbers
in the subscript is the order number of the GHZ-triplet, and the
alphabets inside the bracket, $A$, $B$ and $C$ , represent the
particles within each GHZ-triplet.

\begin{figure}
\begin{center}
\includegraphics[width=9cm,angle=0]{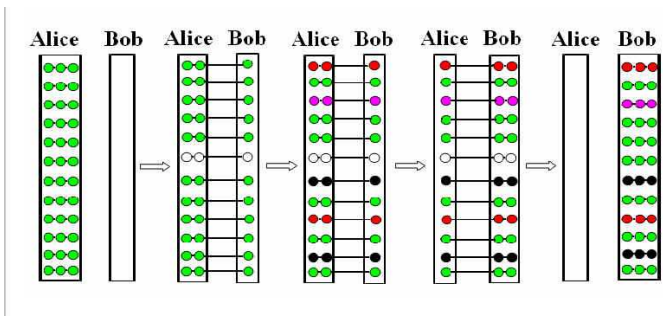}
\caption{ The QSDC procedure using GHZ state.  \label{f1}}
\end{center}
\end{figure}

Step 2: Alice takes the $C$-particle from each GHZ triplet
(particle $P_M(C)$, $M=1,2,...,N$) to form a $C$-sequence,
[$P_1(C)$, $P_2(C)$,... $P_N(C)$]. Then she sends the $C$-sequence
to Bob as shown in Fig \ref{f1}.

Step 3: When Bob receives the $C$-sequence [$P_1(C)$, $P_2(C)$,...
$P_N(C)$]. Alice and Bob perform a security check to see if there is
eavesdropping in the line. Security check can be realized by some
measurements so that the entangled states collapse. Here we provide
two alternatives.

\emph{Method 1 contains the following steps.} First, Bob randomly
chooses particles from his $C$-sequence to make either $\sigma_z$
or $\sigma_x$ measurement. This collapses the GHZ state into
either $|000\ket$ or $|111\ket$ with equal probability if
$\sigma_z$ is measured, or into $(|+x+x\ket+|-x-x\ket)/\sqrt{2}$
if $\vert +x\rangle$ is obtained, $(|+x-x\ket+|-x+x\ket)/\sqrt{2}$
if $\vert -x\rangle$ is obtained when $\sigma_x$ is measured. Then
he announces the positions of these particles and the kind of
measurement he has made. With this information, Alice can make a
$\sigma_z$ measurement on the two particles in the corresponding
GHZ-triplet if Bob measures $\sigma_z$, or a $\sigma_x$
measurement on the two particles in the corresponding GHZ-triplet
if Bob measures $\sigma_x$. They check the results to discover the
existence of Eve. If the channel is safe, the results must be
completely correlated, i.e. if Bob get $|0\rangle(|1\rangle)$,
then Alice get $|00\rangle(|11\rangle)$ when they choose their
measurements along z-direction, and similar case happens for the
$\sigma_x$. For instance, if Eve gets hold the $C$ particle and
sends a fake $C'$ particle to Bob, this security check will
discover her.

\emph{Method 2 contains the following steps.} First, Bob randomly
chooses particles from his $C$-sequence and he also measures each
of these chosen particles using either $\sigma_z$ or $\sigma_x$.
This collapses the GHZ state into either $|000\ket$ or $|111\ket$
with equal probability if $\sigma_z$ is measured, or into
$(|00\ket+|11\ket)/\sqrt{2}$ if $\vert +x\rangle$ is obtained,
$(|00\ket-|11\ket)/\sqrt{2}$ if $\vert -x\rangle$ is obtained when
$\sigma_x$ is measured. Then he announces the positions of these
particles and the kind of measurement he has made. With this
information, Alice can make a $\sigma_z$ measurement on the two
particles in the corresponding GHZ-triplet if Bob measures
$\sigma_z$, or a Bell-basis measurement if Bob measures
$\sigma_x$. They then publish their measured result to check the
existence of Eve. If the channel is safe, the results must be
completely in accordance. On the condition of safe channel, they
go to the next step, otherwise they stop the communication.


Step 4: Alice encodes her information on the remaining two qubits
message sequence, or the $AB$-sequence, [$P_1(A)P_1(B)$,
$P_2(A)P_2(B)$, ..., $P_N(A)P_N(B)$] by performing the unitary
operations, which has been described by equations (\ref{e3}) and
(\ref{e4}). For the secure communication, Alice selects some
sampling pairs of her particles, randomly chosen from  the
$AB$-sequence and  performs one of the eight operations randomly on
them. The remaining pairs of particles in the $AB$-sequence are
encoded with secret messages.  After that she sends the second block
of particles [$P_1(B)$, $P_2(B)$,... $P_N(B)$] which is formed by
the $B$ particles of the GHZ state to Bob. After Bob receives these
$B$ particles, they make a security check in which Alice starts the
security check by choosing some of the sampling pairs for
measurement using one of the checking method described in step 3. If
the security check is passed, Alice then sends Bob the last sequence
[$P_1(A)$, $P_2(A)$,... $P_N(A)$] which consist of the $A$ particles
of the GHZ state. Upon receiving this $A$ particle sequence, they
also perform a security check. If the security check is passed, they
go to the next step.

Step 5: On Bob's side, when he receives the three blocks of
particles step by step, he can make a GHZ state measurement on each
of the GHZ-state and gets the state exactly as shown in equation
(\ref{e2}) and thus gets the message which represent the message
Alice sends.

Step 6: If the $N$ GHZ-states does not complete the task of
communicating the secret message, then they continue with the next
batch of $N$ GHZ-states, starting from step 1.

Under noisy channel, quantum error correction and privacy
amplification have to be performed.

Compared to the two-particle Bell-basis state QSDC protocol, the
three-particle GHZ state can transmit three bits of information each
time though only two encoding unitary operations are performed on
only two qubits, a characteristic signature of the superdense
coding. When multi-partite entangled state are used, it will
transmit more number of bits. The generalization of this scheme to
multi-particle entangled states is given in the following. First
Alice produces a squence of $N$ p-particle entangled states in the
form
\begin{equation}
|\Psi\rangle=\frac{1}{\sqrt{2}}(|0_10_2....0_p\rangle+|1_11_2....1_p\rangle).
\end{equation}
The $p$th particle from each multiplet is sends to Bob first, and
upon receiving this $p$th particle sequence, Alice and Bob perform
security check so that they can assure the security of the
channel. After assuring the security of the channel, Alice
performs encoding operations on the remaining $p-1$ particles in
each multiplet of particles: for the $(p-1)$th particle, four
unitary operations are possible, namely $I$, $\sigma_x$,
$i\sigma_y$ and $\sigma_z$, but for the remaining $p-2$ particles,
only two unitary operations, $\sigma_x$ and $I$ are allowed.
Though the unitary operations performed on only $p-1$ particles,
the number of independent unitary operations is however $4\times
2\times\cdots\times 2=2^p$. Each time , a multiplet of entangled
particles can transmit $p$ bits of classical information.

Another possible generalization of the scheme is to higher
dimensions, namely qubit is generalized into qudit as in Ref.
\cite{long3} in superdense coding. The maximally entangled GHZ-
state in $d$-dimensional Hilbert space is
\begin{equation}
|\Psi_n\rangle=\frac{1}{\sqrt{d}}\sum\limits_{n}|n\rangle_{A}\otimes
|n\rangle_{B}\otimes|n\rangle_{C}
\end{equation}
where $n=0,1,2,...,d-1$. The single particle unitary operations used
for superdense coding are
\begin{equation}
U_{mn} =\sum_{j} e^{2\pi ijm/d}\vert j+n\;{\rm mod} \; d \rangle
\langle j\vert.\label{e8}
\end{equation}
Using these unitary operations $U_{mn}$, one can construct
superdense coding with $d$-dimensional single particle Hilbert
space. To depict it explicitly, we formalize the multi-step
superdense coding scheme in three-level Hilbert space. The
three-level GHZ state is:
\begin{equation}
|\Psi_{00}\rangle_{ABC}=\frac{1}{\sqrt{3}}(|000\rangle_{ABC}+
|111\rangle_{ABC}+|222\rangle_{ABC}).\label{e9}
\end{equation}

By applying single qutrit unitary operations given in (\ref{e8})
with $d=3$ on the first two particles $A$ and $B$ in the following
combination, $U_{00}(A)\otimes U_{mn}(B)$, $U_{01}(A)\otimes
U_{mn}(B)$, $U_{02}(A)\otimes U_{mn}(B)$, where $m,n\in{0,1,2}$, one
can make unique transformation from state (\ref{e9}) to any of the
27 independent three-qudits GHZ-state. For example,
\begin{equation}
\begin{split}
 (U_{00}\otimes
U_{mn})|\Psi_{00}\rangle&=\frac{1}{\sqrt{3}}(|0n0\rangle+e^{2m\pi
i/3}|1 (n+1\;{\rm mod} \;  d) 1\rangle \\ &+e^{4m\pi
i/3}|2 (n+2\;{\rm mod} \;  d)2\rangle),\\
 (U_{01}\otimes
U_{mn})|\Psi_{00}\rangle&=\frac{1}{\sqrt{3}}(|1n0\rangle+e^{2m\pi
i/3}|2 (n+1\;{\rm mod} \;  d) 1\rangle \\ &+e^{4m\pi
i/3}|0 (n+2\;{\rm mod} \;  d)2\rangle),\\
 (U_{02}\otimes
U_{mn})|\Psi_{00}\rangle&=\frac{1}{\sqrt{3}}(|2n0\rangle+e^{2m\pi
i/3}|0 (n+1\;{\rm mod} \;  d) 1\rangle \\ &+e^{4m\pi
i/3}|1 (n+2\;{\rm mod} \;  d) 2\rangle).\\
\end{split}
\end{equation}
Now we come to the multi-step QSDC using qudit superdense coding. In
order to encode the secret message, Alice and Bob make an agreement
on that each of the unitary operations $U_{00}(A)\otimes U_{mn}(B)$,
$U_{01}(A)\otimes U_{mn}(B)$, $U_{02}(A)\otimes U_{mn}(B)$, where
$m,n\in{0,1,2}$, represent one number in the three ternary number
such 000, 012 etc. Then the details of the protocol will be similar
for the qubit GHZ-state case. The number of the possible unitary
operations is $3^p$ which is a direct generalization of that with
two-level quantum system, where $p$ is the number of particles.

It is worth emphasizing that the multi-step should not be replaced
by a three-step process because it sacrifices the security. For
instance, if we divide the $p-1$ particles into two parts with
$p_{1}$ and $p_{2}$ each. First Alice sends Bob the last particle in
each GHZ-state. Then she may perform the encoding operations on the
remaining $p-1$ particles and send the $p_{1}$ first, and then later
send the remaining $p_{2}$ particles after the security of the
$p_{1}$ particles is assured. However, when Eve intercept those
$p_{1}$ particles, she can obtain partial information. In the
original GHZ-state, a measurement of the $p_{1}$ particles yields
either $00\cdots0$ of $11\cdots1$, but after some unitary
transformation, the measurement may obtain either $10\cdots1$ of
$01\cdots0$. Eve may obtain partial information from such
measurement. These partial information contains some secret message
hence is risky. This is in striking contrast to quantum key
distribution there the keys do not carry the secret message, one can
simply drop the raw key if eavesdropping is discovered. By sending
the particles in multi-steps, this danger is avoided since Eve's
measurement on a single particle does not reveal any useful
information.

The security of this multi-step QSDC with GHZ state is similar to
that using EPR entangled states in Ref.\cite{long1}. When the
entangled particles are not in one side, it is just like the BBM92
QKD protocol\cite{bbm92}. As the secret information is encoded in
the whole entangled state, Eve can not get useful information if
she just gets part of the entangled state. When Eve is in the
quantum line, the state of the composite system is
\begin{equation}
|\psi\rangle=\sum\limits_{a,b,c\in(0,1)}|\varepsilon\rangle|a,
b\rangle|c\rangle
\end{equation}
where $|\varepsilon\rangle$ is Eve's state used to probe the
particle. $|a, b\rangle$ and $|c\rangle$ are GHZ states shared by
Alice and Bob after the first step transmission. When she
eavesdrops the $C$ particle sequence, the whole system will be:
\begin{equation}
\begin{split}
|\psi\rangle=&\frac{1}{\sqrt{2}}[|00\rangle(\alpha_1|0\rangle|\varepsilon_{00}\rangle
+\beta_1|1\rangle|\varepsilon_{01}\rangle)\\
&+|11\rangle(\beta_2|1\rangle|\varepsilon_{10}\rangle
+\alpha_2|0\rangle|\varepsilon_{11}\rangle)],
\end{split}
\end{equation}
where $\varepsilon_{00}$, $\varepsilon_{01}$, $\varepsilon_{10}$,
$\varepsilon_{11}$ are Eve's states respectively. We can write out
the probe operator
\begin{equation}
\widehat{E}=\left(%
\begin{array}{cc}
  \alpha_1 & \beta_1 \\
  \beta_2 & \alpha_2 \\
\end{array}%
\right).
\end{equation}
Because $\widehat{E}$ is an unitary operator, the complex number
$\alpha_1$, $\alpha_2$, $\beta_1$, $\beta_2$ satisfy
$|\alpha_1|^2+|\beta_1|^2=|\alpha_2|^2+|\beta_2|^2=1$. The error
rate introduced by Eve is
$\epsilon=|\beta_1|^2=|\beta_2|^2=1-|\alpha_1|^2=1-|\alpha_2|^2$.
The security can thus be similarly analyzed.

The discussions above are based on the ideal conditions. If the
channel noise cannot be omitted, the security problems arise. Eve
may captures some of the particles in the sequence and send the left
to Bob through a better channel. If she intercepts the message
sequence, and does a GHZ-state measurement, she may gets some of the
secret message. So there are probabilities of information leakage.
To avoid the dangerous, we consider a entanglement swapping
strategy. Bob perform the quantum entanglement swapping on the
particles he receives and compare the result with Alice's: if the
particles preserved there, the entanglement swapping succeed. They
measures their particles with either $\sigma_z$ or $\sigma_x$. If
they get the correlation results, the secure channel established and
message sending can be protected.

In conclusion, we proposed the multi-step GHZ state quantum secure
direct communication protocol and  generalized it to the high
dimensional qudit case. The security of the protocol is also
discussed. In ideal condition, the protocol is safe against
eavesdropping. In noisy channel,  quantum privacy amplification
has to be used to obtain security.

The authors wish to thank Drs Jian Wang,Quan Zhang, Lin-mei Liang,
Chao-jing Tang for helpful discussions.

This work is supported the National Fundamental Research Program
Grant No. 001CB309308, China National Natural Science Foundation
Grant No. 60433005,  the Hang-Tian Science Fund, and the Excellent
Young University Teachers' Fund of Education Ministry of China.

\end{document}